\documentclass[conference]{IEEEtran}

\usepackage{epsfig}
\usepackage{listings}
\usepackage{multirow}
\usepackage{colordvi}
\usepackage{url}
\usepackage{xcolor}

\graphicspath{{.}{./Images/}}

\renewcommand{\baselinestretch}{0.91}

\begin{document}

\title{Enabling predictable parallelism in single-GPU systems with persistent CUDA threads}

\author{
	\IEEEauthorblockN{Paolo Burgio}
	\IEEEauthorblockA{University of Modena and Reggio Emilia, Modena, Italy}
	\IEEEauthorblockA{\small{paolo.burgio@unimore.it}}
}

\maketitle

\begin{abstract}
Graphics Processing Unit, or GPUs, have been successfully adopted both for graphic computation in 3D applications, and for general purpose application (GP-GPUs), thank to their tremendous performance-per-watt.
Recently, there is a big interest in adopting them also within automotive and avionic industrial settings, imposing for the first time real-time constraints on the design of such devices.
Unfortunately, it is extremely hard to extract timing guarantees from modern GPU designs, and current approaches rely on a model where the GPU is treated as a unique monolithic execution device.
Unlike state-of-the-art of research, we try to ``open the box'' of modern GPU architectures, providing a clean way to exploit intra-GPU predictable execution.
\end{abstract}

\section{Introduction}
\label{sec:introduction}

In the last decade, increasing demand for low-energy computational power from the embedded world met the tremendous performance-per-watt potential of modern the Graphic-Processing Units (GPUs), opening the doors to the adoption these devices in the new generation of embedded systems.
The NVIDIA Tegra family \cite{tegra_x1} is an example of a GPU-based System-on-Chip (SoC) explicitly designed for smartphones and tablets.
Recently, there is an increasing interest to adopt GPUs also in automotive and avionics settings, imposing for the first time hard real-time constraints in their design.
Unfortunately, GPUs are not tailored to real-time systems, due to the complex hardware structure.
Their aggressively parallel designs extract the maximum performance from the hardware, but at the same time they also hassle the analyzability of the overall platform.
As an addition, the non-openness of most GPU low-level drivers and firmware makes it cumbersome to treat them other than as a ``monolithic'' piece of hardware, which cannot be managed and exploited in a more flexible manner.
A significant example is the \textit{warp scheduler} of NVIDIA GPUs, whose mixed hardware/software structure, firmware and OS drivers represent the key added value of the provider hence they are not disclosed nor well documented.
Although perfectly comprehensible from a business/market point of view, this decision prevents most of the academics world to do research on GPUs, especially in the field of real-time systems.
Indeed, in the real-time domain, a deep comprehension of the device hardware architecture is paramount to achieve predictability/real-time guarantees also at the software level.

Some research has been carried on for supporting timing-accurate and predictable computing on GPUs: for instance, GPUSync \cite{elliott_RTSS13} provides i) pinning mechanism ii) budgeting and iii) integration support for multi-GPU systems.
However, all of the current approaches target multiple-GPU systems, where the device itself is treated as a unique ``atomic'' execution device.
\begin{figure}[!t]
	  \centering
	  \epsfig{file=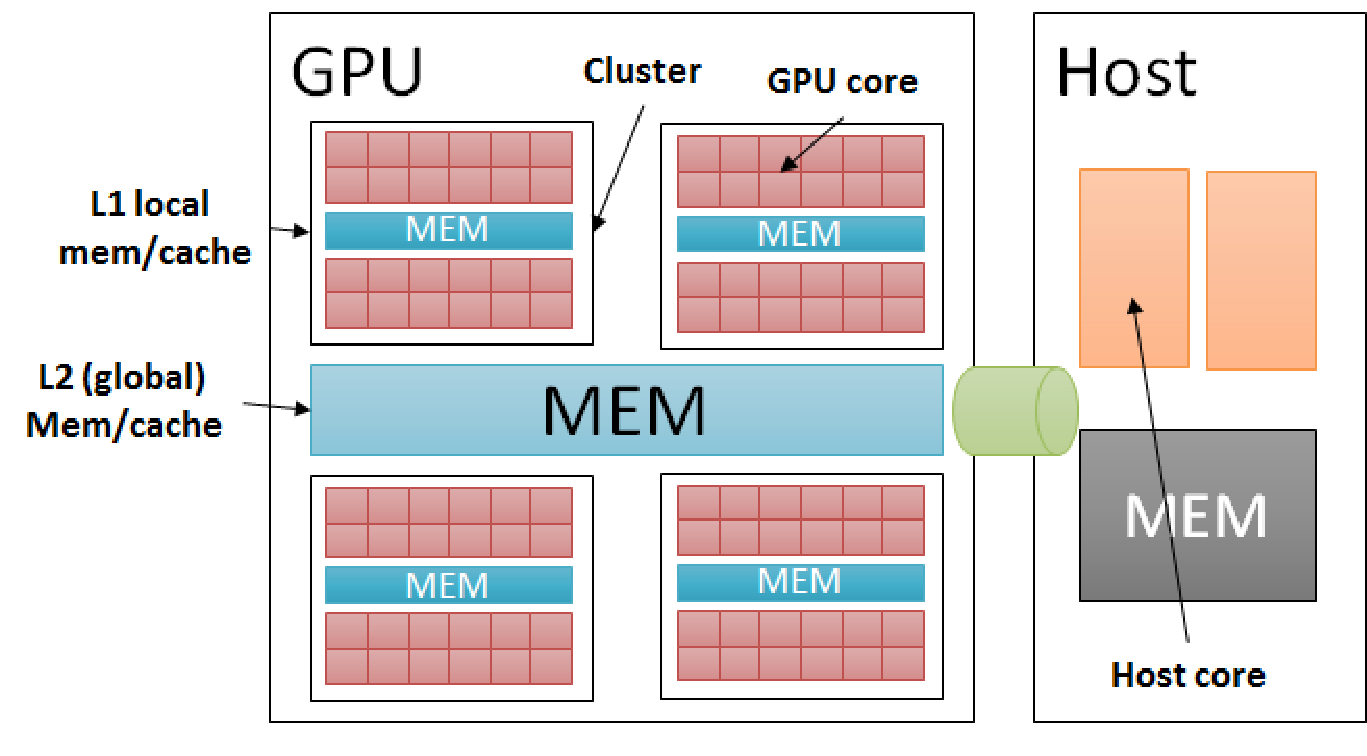, width=.85\columnwidth}
	  \caption{Clustered GPU architecture}
	  \label{fig:gpuarch}
\vspace{-0.4cm}
\end{figure}
This paper described our ongoing effort to ``open the box'' of GPU devices (here specifically of NVIDIA devices, as a testbed), to explore whether it is possible to \textbf{extract predictability guarantees within a single GPU}.
To the best of our knowledge, we are the first doing so, and we will treat the problem \textbf{exclusively from a software point of view}, that is, we will not propose any modifications to existing hardware.

We believe that the generic clustered structure of a GPU device (shown in Figure \ref{fig:gpuarch}) lends itself pretty well to being ``opened'' and treated at a finer-grained architectural level.
We adopt a recent programming paradigm for GPUs which addresses a lightweight and flexible execution model employing \textit{persistent GPU threads} \cite{gupta2012study, capodieci_PAAP15}.
Persistent Threads run at user-level on the GPU, and contrarily to ``traditional'' GPU threads, they are not tailored for a specific computation, but continuously spin-wait for work to execute.
As soon as the host subsystem (e.g., in consumer GPU systems, a multi-core CPU) wants to offload some work to the GPU, it sends to the persistent thread both a descriptor of the work and a reference to the in/out data items, which indicates the actual work to perform.
This scheme can be repeated indefinitely.
With such an approach, it is therefore possible to allocate work on a specific subset of GPU cores in such a way to minimize inter-cores interference (e.g., thrashing of global cache lines) and increasing overall platform predictability.
We are currently developing an implementation named \textbf{LightKernel (LK)}, which we will release as open source software\footnote{Feel free to download this preliminary implementation from:\\ \url{http://hipert.mat.unimore.it/LightKer}} to support research in this direction.

\section{Approach and implementation of LightKernel}
\label{sec:description}

\subsection{Approach and design choices}
Before introducing the design of our LK architecture, it is important to point out a few choices that guided our implementation.
\begin{itemize}
	\item We target real-time systems, where \textit{predictability} is a primary concern. In such systems, it is perfectly normal to trade \textbf{average} performance for \textbf{worst case} performance, and to follow clean, modular software designs, which ensure spatial and timing isolation among application components.
	\item A typical optimization for GPU devices is to share the program counter (PC) register among groups of hardware cores. This causes the so-called \textit{lockstep} execution, which means that cores belonging to the same group are tied to execute the very same assembly instruction, hence their execution cannot diverge. This mechanism is extremely useful when implementing data-parallel execution pattern.
	\item In modern GPUs, computing cores are physically partitioned into clusters (see Figure \ref{fig:gpuarch}), with local (L1) caches and software-managed memory banks to maximize data locality to computation. More ``lockstep groups'' can belong to the same cluster. In such a system, exploiting local memories to maximize the locality of data to computing elements is a must to achieve performance. For this reason, we decide to expose clusters at the application level, providing low-level software subroutines to map both data and work on a specific cluster.
\end{itemize}

\subsection{NVIDIA terminology and CUDA}
Our implementation is based on CUDA \cite{cudaprogrammingguide}, which is NVIDIA's proprietary API for programming GPUs.
We chose these devices as a testbed because NVIDIA is undoubtedly the GPU market leader, and CUDA is one of the most widely adopted programming model for GPUs.
Our approach can be seamlessly ported, e.g., to OpenCL \cite{opencl} with minimal effort.
In the NVIDIA terminology, the GPU device is composed of \textit{CUDA cores}, clustered onto \textit{Streaming Multiprocessors -- SMs}.
CUDA programmers partition the application onto a two-dimensional work space composed of \textit{CUDA threads}, grouped onto \textit{thread blocks}.
The portion(s of application that are offloaded to the GPU device are called \textit{CUDA kernels}.
Within a kernel, typically, threads are mapped onto CUDA cores, while thread blocks are mapped onto GPU SMs.
Lockstep islands are called \textit{warps}, composed by multiple threads executing in ``Single Program Multiple Data'' -- SPMD fashion.
Intuitively, one thread block is composed of multiple warps, whose dimension is fixed for a given GPU architecture, and which are directly managed by the CUDA runtime.
From the real-time point of view, warps are the data-parallel, non-preemptable, atomic unit of work which is schedulable on the GPU device, and programmers have no visibility nor control over them.
We believe this is the biggest limitation of GPU architectures, from a real-time point of view.

\subsection{LK persistent kernel}
In a first step we follow the simplest design possible, and decide to statically \textit{pin} persistent threads on GPU clusters at boot time, and to dynamically allocate work to a specific thread, that is, to a specific cluster of cores.
This potentially enables fine-grained execution models, exploiting intra-GPU parallelism in a flexible way.
This is possible because, as explained, ``hardware lockstep islands'' are confined within the single clusters.
Roughly speaking, we spawn a single (persistent) CUDA kernel, made of \textit{B} blocks of \textit{T} CUDA threads, where \textit{B} is the number of SMs in the target device, and \textit{T} is the number of CUDA cores within a single SM, which is fixed for a given GPU architecture.
What typically happens in GPU runtimes is that, when a CUDA kernel is spawned, its blocks are assigned to SMs in a round-robin fashion, so each LK block is mapped onto a dedicated SMs.
We implemented a low-level checking mechanism for this (each thread within a block reads the SM number and compares it with the block ID which is assigned by CUDA runtime), which however was never triggered during our development and experiments.

Even if in modern GPUs it is possible to create more CUDA blocks (resp. threads) than available SMs (resp. cores), we want our design to be simple, and we don't spawn more threads than cores, and map exactly one CUDA block onto one SM \footnote{Architectural features such as, e.g., number of CUDA cores and SMs in a given GPUs can be easily obtained using platform-specific hooks.}.
We will explore more complex execution models, based on multiple persistent threads, in a future step.
\begin{figure}[ht]
	  \centering
	  \epsfig{file=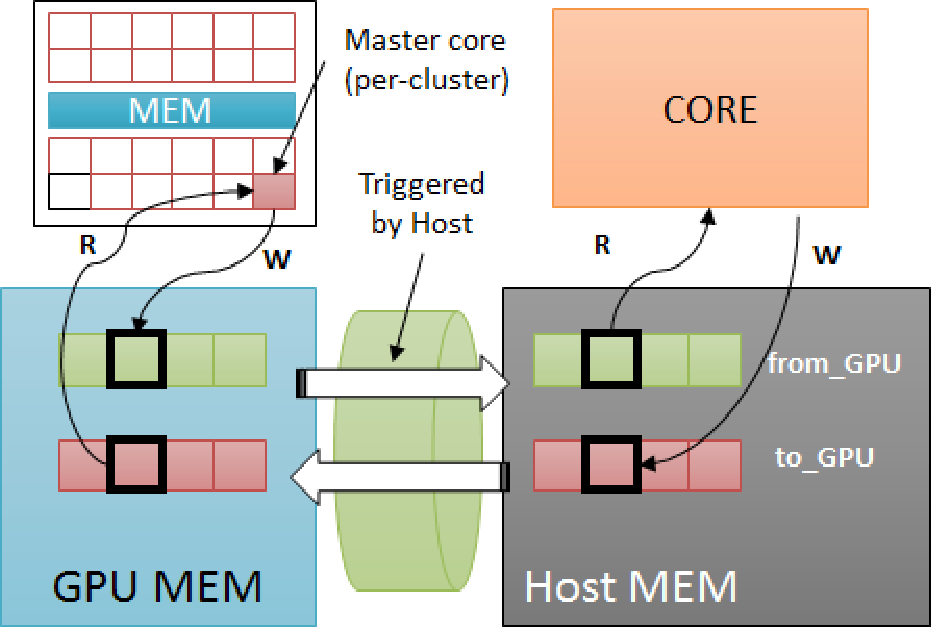, width=.7\columnwidth}
	  \caption{Dual mailbox structure}
	  \label{fig:mailbox}
\end{figure}
\subsection{Host-to-device communication}
A key aspect of GPU computing is is the synchronization between the host and the device.
We want to deliver work at the granularity of the single SM, and to do so, we implemented a dual lock-free mailbox system (see for instance \cite{marongiu_MICPRO11}).
In such a mechanism, each SM/persistent thread has two dedicated mailbox items (called \texttt{from\_GPU} and \texttt{to\_GPU}), as shown in Figure \ref{fig:mailbox}.
We promote one CUDA thread within each block to act as \textit{master thread} and to read-write the corresponding two mailboxes.
Mailboxes are implemented as C \texttt{int}egers, whose values represent the statuses of LK-to-host communication protocol, as in Table \ref{tab:mailbox}.
\begin{table}[!t]
\begin{center} \small
\begin{tabular}{|c|c||c|c|}
\hline
Persistent thrd status	& Value		& Persistent thrd status	& Value		\\	\hline
\textit{from\_GPU}			&  				& \textit{to\_GPU}				&  				\\	\hline
THREAD\_INIT						& 0				& THREAD\_NOP							& 4				\\	\hline
THREAD\_FINISHED				& 1 			& THREAD\_EXIT						& 8 			\\	\hline
THREAD\_WORKING					& 2 			& THREAD\_WORK						& 16+ 	  \\	\hline
THREAD\_NOP							& 4				&													&					\\	\hline
\end{tabular}
\caption{Persistent thread statuses and mailbox values}
\label{tab:mailbox}
\vspace{-0.8cm}
\end{center}
\end{table}
In the considered architecture (see Section \ref{sec:validation}) there are 16 SMs/clusters, hence our mailbox is composed by two arrays of 4x16 = 64 bits.
Typically, GPUs are connected to CPU hosts via PCI EXPRESS connectors, delivering more than 15 GByte/sec.
However, this is \textbf{peak} performance, i.e., it is delivered only with big data chunks, while the size of our mailbox is too narrow to efficiently exploit the mechanism, and we experience performance degradation in some cases.
Unfortunately, data transfer from and to the device is mastered by OS driver for the specific device, which is proprietary and undisclosed, so there is not much chance to improve this mechanism without accessing its internals.

\section{Preliminary evaluation campaign}
\label{sec:validation}
We performed a set of experiments to compare our LK to ``standard'' CUDA kernels.
We expect the mailbox mechanism to be much lighter of traditional kernel spawn, with a significant performance gain.
The only point of concern, as introduced in Section \ref{sec:description}, is the efficiency of transmitting few data chunks (i.e., the mailbox) across the PCI connector.
We performed the experiments on a machine with the state-of-the-art of consumer GPUs by NVIDIA, the GTX980 mode, with the host running Ubuntu 14 Linux on an Intel i7 quad-core with 8GB RAM clocked at 3.6 GHz.

To structure the experiments, we split LK execution in a \textit{Init} phase, to boot the system, and a \textit{Dispose} phase, to release the GPU resource.
In the middle, following the persistent thread model, we perform multiple subsequent stages to offload work to the device.
Each of this stages is composed of four phases, namely \textit{Copyin} of data in the device memory, \textit{Trigger} of one or multiple SMs, \textit{Wait} for one or multiple SM, and \textit{Copyout} of data to the host memory.
Similarly, traditional CUDA kernels, are made of an \textit{Alloc} phase, to initialize the device driver and data buffers, \textit{Copyin}, \textit{Launch}, \textit{Wait}, \textit{Copyout}, and \textit{Dispose} phase.
We currently will not focus on data transfer effect, i.e., we do not consider the Copyin and Copyout phases, which are strongly data/application dependent.
We implemented a simple benchmark which performs a loop of 20k iterations before exiting, representative of a ``medium'' size GPU kernel, and is completely computation bound (no memory transfer or accesses).

We performed 100 experiments for each configuration, both for LK and traditional CUDA.
We performed two sets of experiments: one where only one GPU core is used, and one where the full GPU is used.
Table \ref{tab:avg1} shows the time (clock cycles spent on the host) to perform the aforementioned phases.
\begin{table}[ht]
\begin{center} \small
\begin{tabular}{|c|c|c|c|}
\hline
\multicolumn{4}{|c|}{\textbf{Single SM}}											\\	\hline
LK Init				& LK Trigger		& LK Wait			& LK Dispose			\\	\hline
509M					& 239						& 190k				& 30M							\\	\hline
CUDA Alloc		& CUDA Spawn    & CUDA Wait		& CUDA Dispose		\\	\hline
496M					& 3.9k					& 175k				& 274k						\\	\hline
\multicolumn{4}{|c|}{\textbf{Full GPU}}												\\	\hline
LK Init				& LK Trigger		& LK Wait			& LK Dispose			\\	\hline
503M					& 210						& 190k				& 30M							\\	\hline
CUDA Alloc		& CUDA Spawn    & CUDA Wait		& CUDA Dispose		\\	\hline
497M					& 3.8k					& 176k				& 247k						\\	\hline
\end{tabular}
\caption{Average values for LK and traditional CUDA (single SM)}
\label{tab:avg1}
\vspace{-0.2cm}
\end{center}
\end{table}
%
%

Unfortunately, while performing the former experiment, that is, with a single-SM, in some cases the GPU device got stuck.
This is due to the fact that triggering a single SM means transfering only a few bytes of data across the PCI connector (the associated mailbox items), and in some cases the optimization mechanism at the driver-level indefinitely postpones this (excessively small) transfer.
We therefore were forced to transfer the full mailbox also in this cases, and this explains how the numbers in the two tables are comparable.
Nicely, we see that LK outperforms ``standard'' CUDA by a factor of 10$\times $ for the Trigger phase.
This means that we are more ``reactive'', and potentially capable of handling finer-grained kernels (on the order of few thousands of clock cycles), because the overhead to offload work on the GPU is smaller.
This is a promising result also for non-real-time GPU computing \textit{tout court}.
Unfortunately, on the other hand, the Wait phases behaves similarly (around 170k vs. 190k cycles) for LK and CUDA, as explained, because of low-level data transfer policies by the driver and OS.
Further optimizations are needed at this point, and we plan to do them as a next step of the project.
Init (resp. Alloc) and Dispose phase are less interesting from a point of view of LK, because they are only performed at system boot and shutdown.
We anyhow notice how the latter phase is one order of magnitude slower in LK.

\begin{table}[ht]
\begin{center} \small
\begin{tabular}{|c|c|c|c|}
\hline
LK Init				& LK Trigger		& LK Wait			& LK Dispose			\\	\hline
521M					& 1.1k					& 203k				& 33M							\\	\hline
CUDA Alloc		& CUDA Spawn    & CUDA Wait		& CUDA Dispose		\\	\hline
501M					& 7.7k					& 176k				& 893k							\\	\hline
\end{tabular}
\caption{Worst values for LK and traditional CUDA (single SM))}
\label{tab:worst}
\vspace{-0.2cm}
\end{center}
\end{table}
Table \ref{tab:worst} shows the worst case times for the considered phases, only for the case of single SM (numbers involving the full GPU are similar).
We see that both for LK and traditional CUDA, the variance is significant for the Trigger phase, while LK also suffers some variance against average performance for the Wait phase.
This variance in platform performance is crucial in the real-time domain, where worst case performance (and its difference with average-case performance) must be minimized.
For this reason, we will explore on low-level software optimization as a next step of the project.

\section{Conclusion and Acknowledgement}
\label{sec:conclusion}
GPUs are extremely powerful machines, but they are not yet ready for adoption within industrial real-time settings.
This is mainly due to their complex architecture, and non-openness of software runtime and drivers.
As opposed to state-of-the-art of real-time GPU computing, we exploit intra-GPU parallelism, and provide a framework to explore real-time capabilities of most advanced architectures within the single GPU devices.
The framework is not yet completed, but current results are promising, and we already identified research and development paths for our \textsl{LightKernel} tool.

This work was supported by the HERCULES Project, funded by European Union's Horizon 2020 research and innovation program under grant agreement No. 688860


\bibliographystyle{IEEEtran}
\bibliography{biblio_CR}

\end{document}